\documentclass[twocolumn,showpacs,preprintnumbers,amsmath,amssymb,prb]{revtex4}

\usepackage{graphicx}% Include figure files
\usepackage{dcolumn}% Align table columns on decimal point
\usepackage{bm}% bold math
\usepackage{units}%
\usepackage[breaklinks=true,pdfborder={0 0 0}]{hyperref}

\usepackage{tabularx}

\usepackage{color}
\usepackage{tikz}

\begin{document}
\newlength{\LL} \LL 1\linewidth

\title{Insight into the skew-scattering mechanism of the spin Hall effect:\\
potential scattering versus spin-orbit scattering}
% Force line breaks with \\

\author{Christian Herschbach}
\email{cherschb@mpi-halle.mpg.de}
\affiliation{Max Planck Institute of Microstructure Physics, Weinberg 2,
06120 Halle, Germany}
\affiliation{Institute of Physics, Martin Luther University Halle-Wittenberg,
06099 Halle, Germany}
\author{Dmitry V. Fedorov}
\affiliation{Max Planck Institute of Microstructure Physics, Weinberg 2,
06120 Halle, Germany}
\affiliation{Institute of Physics, Martin Luther University Halle-Wittenberg,
06099 Halle, Germany}
\author{Ingrid Mertig}
\affiliation{Max Planck Institute of Microstructure Physics, Weinberg 2,
06120 Halle, Germany}
\affiliation{Institute of Physics, Martin Luther University Halle-Wittenberg,
06099 Halle, Germany}

\author{Martin Gradhand}
\affiliation{H.~H.~Wills Physics Laboratory,
University of Bristol, Bristol BS8 1TL, United Kingdom}

\author{Kristina Chadova}
\affiliation{Department of Chemistry, Physical Chemistry,
Ludwig-Maximilians University Munich, Germany}
\author{Hubert Ebert}
\affiliation{Department of Chemistry, Physical Chemistry,
Ludwig-Maximilians University Munich, Germany}
\author{Diemo K\"odderitzsch}
\affiliation{Department of Chemistry, Physical Chemistry,
Ludwig-Maximilians University Munich, Germany}

\date{\today}% It is always \today, today,

\begin{abstract}
We present a detailed analysis of the skew-scattering contribution to the spin Hall
conductivity using an extended version of the resonant scattering model of Fert
and Levy [Phys. Rev. Lett. {\bf 106}, 157208 (2011)]. For $5d$ impurities in a Cu host,
the proposed phase shift model reproduces the corresponding
first-principles calculations. Crucial for that agreement is the consideration of two
scattering channels related to $p$ and $d$ impurity states, since the discussed
mechanism is governed by a subtle interplay between the spin-orbit and potential
scattering in both angular-momentum channels. It is shown that the potential
scattering strength plays a decisive role for the magnitude of the spin Hall
conductivity.
\end{abstract}

%Use showkeys class option if keyword
\maketitle
An intriguing direction in the development of spintronic devices is based on the spin Hall
effect (SHE).~\cite{Dyakonov71_Hirsch99} This phenomenon, caused by spin-orbit coupling (SOC),
provides an opportunity for spin current generation in nonmagnetic materials without injection
from ferromagnets. Materials with a large spin Hall angle (SHA), describing the efficiency
of charge into spin current conversion, are highly desirable. Recently, a number of materials
with a \emph{giant} SHE, corresponding to SHA's of the order of $0.1$, were
predicted~\cite{Tanaka08,Gradhand10} and observed experimentally.~\cite{Seki08,Liu12,Niimi12}
Some of these studies~\cite{Gradhand10,SPIN,Niimi12} indicate that tuning the skew-scattering
mechanism by an appropriate choice of impurities, especially in noble metals, is a promising
route to obtain a giant SHE.

For the understanding of the essential conditions of such a strong SHE, a microscopic analysis of this mechanism
is desired. In particular, models with input parameters provided by first-principles calculations are helpful
for an intuitive picture and a detailed analysis of the material specific SHE. Recently, Fert and Levy~\cite{Fert11} have
proposed a resonant scattering model and applied it to a Cu host with $5d$ impurities. The well-known behavior
of the residual resistivity of noble metals with transition metal impurities is qualitatively well described by Friedel's
$d$ resonance model.~\cite{Friedel58,Daniel65,Braspenning82,Mertig_book} Therefore, it was assumed that the spin-orbit
driven transverse transport in these systems is mainly caused by impurity $d$ states.~\cite{Fert11}
In this paper we show that despite such a seemingly sensible assumption, the contribution
related to the spin-orbit scattering in the $p$ channel is comparably large for the considered systems.
Consequently, the total spin Hall conductivity (SHC) is predominantly caused by two scattering
channels related to both $p$ and $d$ states. This somewhat surprising result is caused by the vertex corrections,
which play an ultimate role for the skew-scattering mechanism. They enter the semiclassical approach via the so-called
\emph{scattering-in} term of the Boltzmann equation.~\cite{Gradhand10,Butler85,Lowitzer11} Furthermore, we show that
for the considered $5d$ series of impurity atoms in copper the magnitude of the SHC is mainly determined by
the potential scattering strength reflected in the momentum relaxation time.

The paper is organized in three parts. First, we provide a derivation of the proposed
phase shift model as an extended version of the resonant scattering model considered in
Ref.~\onlinecite{Fert11} for a description of the SHE. Then, we present the results for $5d$
impurities in a Cu host obtained within this model in comparison to \emph{ab initio} results,
to confirm its validity. Finally, we perform a detailed analysis identifying the significant
contributions to the SHC to elucidate our findings outlined above.

Within the semiclassical approach the conductivity tensor
can be written, using the spherical band approximation, as~\cite{Fedorov13}
\begin{equation}\label{eq.:sigma}
\begin{array}{ll}
\hat{\sigma} = \frac{e^2}V \sum\limits_\mathbf{k} \delta ({\cal E}_\mathbf{k} - {\cal E}_F)
\mathbf{v}_\mathbf{k} \circ \boldsymbol\Lambda_\mathbf{k} =
\frac{e^2 m_e k_{\rm F}}{\hbar^2 (2\pi)^3} \int {\rm d}
\Omega_\mathbf{k}~\mathbf{v}_\mathbf{k} \circ \boldsymbol\Lambda_\mathbf{k}\ ,
\end{array}
\end{equation}
where the mean free path is given by the Boltzmann equation~\cite{Mertig99}
\begin{equation}\label{eq.:Lambda}
\begin{array}{ll}
\boldsymbol\Lambda_\mathbf{k} = \boldsymbol\Lambda_\mathbf{k}^{\rm out} + \boldsymbol\Lambda_\mathbf{k}^{\rm in} =
\tau_\mathbf{k} ( \mathbf{v}_\mathbf{k} + \sum\limits_{\bf k'}
P_{\mathbf{k} \gets \mathbf{k'}} \boldsymbol\Lambda_\mathbf{k'} )
\end{array}
\end{equation}
with the momentum relaxation time
\begin{equation}\label{eq.:tau_definition}
\begin{array}{ll}
\frac 1{\tau_\mathbf{k}} = \sum\limits_{\bf k'} P_{\mathbf{k'} \gets \mathbf{k}} =
\frac{2\pi}{\hbar} c_i N \sum\limits_{\bf k'}
|T_{\mathbf{k'} \gets \mathbf{k}}|^2 \delta ({\cal E}_\mathbf{k} - {\cal E}_\mathbf{k'})
\end{array}
\end{equation}
and the group velocity $\mathbf{v}_\mathbf{k} = \hbar \mathbf{k} / m_e$. Here, the microscopic transition probability
$P_{\mathbf{k'} \gets \mathbf{k}}$ is the scattering rate from an initial state $\mathbf{k}$ into a final
state $\mathbf{k'}$. This quantity is defined by the corresponding transition matrix $T_{\mathbf{k'} \gets \mathbf{k}}$
and scales with the impurity concentration $c_i = N_i / N$, as valid for noninteracting impurities.~\cite{Mertig99}

Of crucial importance for the further discussion of the considered phenomenon is the second term on the r.h.s. of
Eq.~(\ref{eq.:Lambda}). This scattering-in term, which corresponds to the vertex corrections of the Kubo theory
in the dilute limit of impurity concentrations,~\cite{Butler85} is known to be responsible for the skew-scattering
mechanism.~\cite{Gradhand10,Fert11,Lowitzer11,Sinitsyn08} Moreover, only the antisymmetric part of 
$P_{\mathbf{k} \gets \mathbf{k'}}$ involved in Eq.~(\ref{eq.:Lambda}) contributes to the effect.~\cite{Fert11,Fedorov13}

Following Refs.~\onlinecite{Fert11} and \onlinecite{Fert81}, the transition matrix for the ``spin-up'' (``$+$'') states
can be written, to first order in $\lambda_l / \Delta_l$, as
\begin{equation}\label{Tkk}
\begin{array}{ll}
T_{{\bf k'} \gets {\bf k}}^{+ \gets +} = \frac{4 \pi^2 \hbar^2}{m_e k_{\rm F} V} \sum\limits_{lm}
[ m \frac{\lambda_l}{\Delta_l} e^{{\rm i} 2 \eta_l}
\sin^2{\eta_l}  - 2 e^{{\rm i} \eta_l} \sin{\eta_l} ] \\ \qquad\qquad\qquad\qquad\qquad\qquad
\times \left(Y_l^m ({\bf \hat{k}})\right)^* Y_l^m ({\bf \hat{k'}})\ .
\end{array}
\end{equation}
Here, $\Delta_l$ and $\lambda_l$ are the resonance width and the SOC constant for the corresponding $l>0$ impurity
level, respectively. This expression can be derived starting from the conventional nonrelativistic transition
matrix and assuming the phase shifts to become $m$-dependent as
$\eta_l \to \eta_l^m \approx \eta_l - \frac{m \lambda_l}{2 \Delta_l}\sin^2{\eta_l}$ for a perturbative
treatment of the SOC.~\cite{Fert81,Fert72} Thus, in Eq.~(\ref{Tkk}) the spin-orbit scattering strength is measured by
$(\lambda_l / \Delta_l)\sin^2{\eta_l}$, while $\sin{\eta_l}$ in the second term relates to the strength of
the potential scattering.~\cite{Comment}

As mentioned above, the skew-scattering mechanism is solely determined by the antisymmetric part of the microscopic
transition probability. Thus, one needs to calculate
$|T_{{\bf k'} \gets {\bf k}}|_{\rm antisym}^2 = \left( |T_{{\bf k'} \gets {\bf k}}|^2 - |T_{{\bf k} \gets {\bf k'}}|^2
\right)/2$. Since the first term on the r.h.s. of Eq.~(\ref{Tkk}) is antisymmetric while
the second one is symmetric with respect to exchange of $\mathbf{k}$ and $\mathbf{k'}$,~\cite{Fert11} we obtain
\begin{equation}\label{Tkk2_asym}
\begin{array}{ll}
|T_{{\bf k'} \gets {\bf k}}^{+ \gets +}|_{\rm antisym}^2 = \frac{64 \pi^4 \hbar^4}{V^2 m_e^2 k_{\rm F}^2} {\rm i} \sum\limits_{lm}
\sum\limits_{l' m'} m \frac{\lambda_l}{\Delta_l} \sin{(2\eta_l-\eta_{l'})} \\ \times \sin^2{\eta_l} \sin{\eta_{l'}}
Y_l^m ({\bf \hat{k}}) \left(Y_l^m ({\bf \hat{k'}})\right)^* \left(Y_{l'}^{m'} ({\bf \hat{k}})\right)^* Y_{l'}^{m'}
({\bf \hat{k'}})\ .
\end{array}
\end{equation}
Within the spherical band approximation, the skew-scattering contribution to the spin Hall conductivity,
$\sigma_{yx}^s = 2 \sigma_{yx}^+$, is given by~\cite{Fedorov13}
\begin{equation}\label{eq.:sigma_skew}
\begin{array}{ll}
\sigma_{yx}^+ = \frac{c_i N V e^2 k_{\rm F}^2 \tau_0^2}{\hbar^3 (2\pi)^5}
\int {\rm d} \Omega_\mathbf{k} \int {\rm d} \Omega_\mathbf{k'}\ k_y k_x^\prime
|T_{\mathbf{k} \gets \mathbf{k'}}^{+ +}|_{\rm antisym}^2\ .
\end{array}
\end{equation}
This is obtained assuming 
$\boldsymbol\Lambda_\mathbf{k'} \approx \boldsymbol\Lambda_\mathbf{k'}^{\rm out} = \tau_\mathbf{k'} \mathbf{v}_\mathbf{k'}$
for the scattering-in term of Eq.~(\ref{eq.:Lambda}) as well as an isotropic momentum relaxation time
$\tau_\mathbf{k} \approx \tau_0$. The involved components of the crystal momentum can be expressed
in terms of spherical harmonics as~\cite{Varshalovich88}
\begin{equation}\label{eq.:Kx_Ky}
\begin{array}{ll}
k_x = \frac{k_{\rm F} \sqrt{2\pi} \left[ Y_1^{-1} ({\bf \hat{k}}) - Y_1^1 ({\bf \hat{k}}) \right]}{\sqrt{3}} ,\ 
k_y = \frac{i k_{\rm F} \sqrt{2\pi}  \left[ Y_1^{-1} ({\bf \hat{k}}) + Y_1^1 ({\bf \hat{k}}) \right]}{\sqrt{3}}\ .
\end{array}
\end{equation}
Taking into account Eq.~(\ref{Tkk2_asym}) and the integrals over three spherical harmonics~\cite{Fert87}, which are
related to the Clebsch-Gordan coefficients,~\cite{Varshalovich88} we obtain~\cite{Supplementary}
\begin{equation}\label{eq.:SHC_final}
\begin{array}{ll}
\sigma_{yx}^+ = \left(\frac{e^2}{\hbar}\right)
\left(\frac{2 \hbar^2 k_{\rm F}^2 c_i}{\pi m_e^2 V_0}\right)\tau_0^2 \times \\
\left\{\frac 13 \frac{\lambda_1 \sin^2{\eta_1}}{\Delta_1}
\left[ \sin{(2\eta_1-\eta_0)} \sin{\eta_0} - \sin{(2\eta_1-\eta_2)} \sin{\eta_2} \right]
\right. \\ \left. + \frac{\lambda_2 \sin^2{\eta_2}}{\Delta_2}
\left[ \sin{(2\eta_2-\eta_1)} \sin{\eta_1} - \sin{(2\eta_2-\eta_3)} \sin{\eta_3} \right] \right. \\ \left.
+ 2 \frac{\lambda_3 \sin^2{\eta_3}}{\Delta_3} \sin{(2\eta_3-\eta_2)} \sin{\eta_2} \right\}\ ,
\end{array}
\end{equation}
where $V_0$ is the unit cell volume. Here, all scattering contributions involving $s$, $p$, $d$, and $f$ states
are considered, neglecting terms with $l > 3$. A more general expression for $\sigma_{yx}^+$ is provided in
the Supplemental Material.~\cite{Supplementary} In addition, for the isotropic momentum relaxation time involved
in Eq.~(\ref{eq.:SHC_final}) we neglect the influence of the SOC and apply the expression~\cite{Mertig_book}
\begin{equation}
\begin{array}{ll}\label{eq.:tau}
\frac 1{\tau_0} = \frac{4\pi\hbar c_i}{m_e k_{\rm F} V_0} \sum\limits_l (2l+1) \sin^2{\eta_l}\ \ ,
\end{array}
\end{equation}
which is similar to the approach used in Ref.~\onlinecite{Fert11}.

For practical applications of Eq.~(\ref{eq.:SHC_final}) one needs to know the SOC constant for an impurity
atom in the considered host. Often this is approximated by the corresponding atomic SOC constant.~\cite{Fert11,Monod82}
However, $\lambda_l$ for an impurity embedded in a crystal can differ significantly from the value of an isolated
atom.~\cite{Yafet63,Fedorov08} Therefore, it is preferable to calculate $\lambda_l$ via the self-consistent electronic
structure of a real impurity system.~\cite{Fedorov08} The resonance width required additionally can be obtained
from the partial local density of states at the impurity site. Of course, the assumption of a simple Lorentz shape
for all impurity states with $l>0$, as used in Eq.~(\ref{Tkk}), is commonly difficult.~\cite{Daniel65} However, we can
generalize Eq.~(\ref{eq.:SHC_final}) using the following relations~\cite{Yafet68}
\begin{equation}\label{eq.:Input}
\begin{array}{ll}
\frac{\lambda_l \sin^2{\eta_l}}{\Delta_l} = \frac {2(\delta_{l-1/2} - \delta_{l+1/2})}{2l+1}\ \ ,\ \ \ \ 
\eta_l = \frac{l\delta_{l-1/2}+(l+1)\delta_{l+1/2}}{2l+1}
\end{array}
\end{equation}
based on the phase shifts related to the relativistic quantum number $j = l \pm \nicefrac 12$. This makes it possible
to apply Eq.~(\ref{eq.:SHC_final}) for any impurity system, since no special assumptions for the impurity states
are made. Nevertheless, the virtual bound states are treated by this generalized approach as well.

For our study, the corresponding relativistic phase shifts $\delta_{l \pm \nicefrac 12}$ are obtained
by first-principles calculations performed with a relativistic Korringa-Kohn-Rostoker Green's function
method.~\cite{Gradhand09} Their values for the considered $5d$ impurities in a Cu host are presented in
the Supplemental Material.~\cite{Supplementary}

\begin{figure}[t!]
\includegraphics[width=0.90\LL]{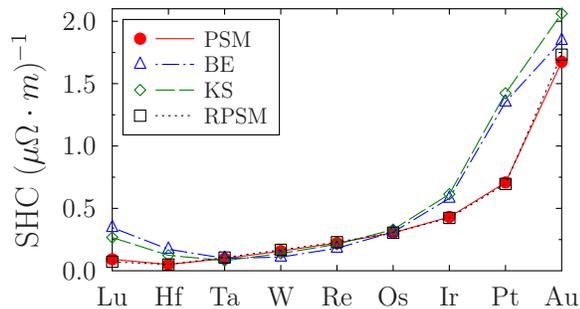}
\caption{(Color online) The spin Hall conductivity for $5d$ impurities in a Cu host obtained by
the considered phase shift model (PSM), the Boltzmann equation (BE), and the Kubo-St\v{r}eda formula (KS).
For comparison, the results obtained within the relativistic phase shift model (RPSM) are shown.}
\label{fig.:SigmaYX_total}
\end{figure}

Figure 1 shows results for the SHC obtained using Eqs.~(\ref{eq.:SHC_final})--(\ref{eq.:Input}) in comparison to
direct first-principles calculations performed applying the Boltzmann equation~\cite{Gradhand10} and the Kubo-St\v{r}eda
formula.~\cite{Lowitzer11} Also included are the corresponding results of the relativistic phase shift model, which
was presented in Ref.~\onlinecite{Fedorov13}. Throughout the paper, all values of the SHC are shown for the impurity
concentration of 1~at.\%. The results of the considered and relativistic phase shift model almost coincide, which
points to an efficient treatment of the SOC within the used perturbative approach. In addition, the models are in
good agreement with the results of the \emph{ab initio} calculations. This is a consequence of the almost spherical
Fermi surface of copper.

\begin{figure}[t]
\includegraphics[width=0.95\LL]{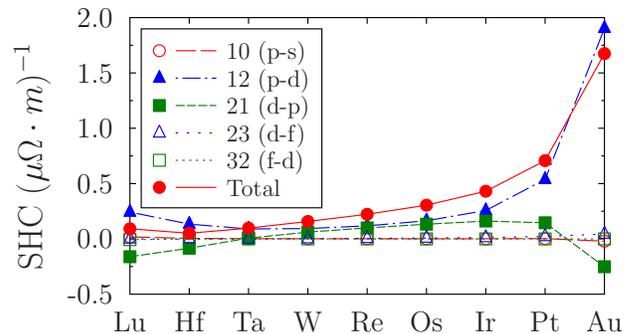}
\caption{(Color online) The spin Hall conductivity for $5d$ impurities in a Cu host obtained by the considered
phase shift model. The different contributions to $\sigma_{yx}^s$ labeled as $l l'$ correspond to $C_{l l'}$
defined by Eq.~(\ref{eq.:C_ll}).}
\label{fig.:SigmaYX_ll}
\end{figure}

In comparison to the first-principles calculations, the models give easy access to a detailed analysis
which helps to identify the most important contributions. The presented model is particularly useful in that
respect, since Eq.~(\ref{eq.:SHC_final}) makes it possible to separate different $l$ channels for the spin-orbit
scattering contributing to the SHC. They are shown in Fig.~\ref{fig.:SigmaYX_ll} and labeled with $l l'$ according
to the related terms
\begin{equation}\label{eq.:C_ll}
\begin{array}{ll}
C_{l l'} = \frac{\lambda_l \sin^2{\eta_l}}{\Delta_l}
\sin{(2 \eta_l - \eta_{l'})} \sin{\eta_{l'}}
\end{array}
\end{equation}
in the braces of Eq.~(\ref{eq.:SHC_final}). Obviously, all contributions involving $s$ and $f$ states are negligible
and the total SHC is provided just by two terms labeled as $p$$-$$d$ and $d$$-$$p$. Surprisingly, the $p$$-$$d$ contribution,
related to the spin-orbit scattering in the $p$ channel, is of the same order as the $d$$-$$p$ contribution. Moreover,
for Pt and Au impurities the $p$$-$$d$ contribution becomes even significantly larger than the $d$$-$$p$ one.

\begin{figure}[t]
\includegraphics[width=0.85\LL]{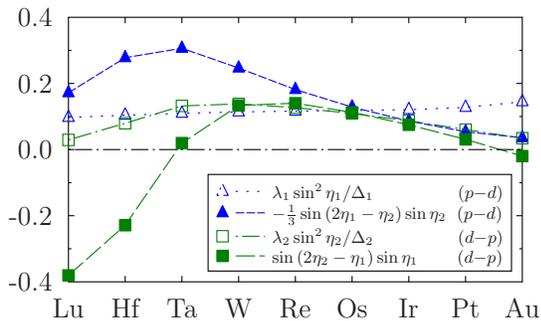}
\caption{(Color online) The two parts of the contributions 
$C_{p-d}$ and $C_{d-p}$ are shown separately.}
\label{fig.:SOC}
\end{figure}

To get a clear insight into the origin of this unexpected result, it is worth to use another helpful property
of Eq.~(\ref{eq.:SHC_final}). The Hall conductivity given by this expression is a result of the interplay between
the spin-orbit and potential scattering, which were entering additively into the transition matrix only.
Nevertheless, the specific structure of Eq.~(\ref{eq.:SHC_final}) still allows one to separate out the spin-orbit
scattering strength from the rest of the skew-scattering contribution to the SHC. In other words, the terms
$C_{p-d} \equiv -\frac 13 C_{1 2}$ and $C_{d-p} \equiv C_{2 1}$ providing the dominant contributions to
Eq.~(\ref{eq.:SHC_final}) can be divided into two factors. The first one, $(\lambda_l / \Delta_l) \sin^2{\eta_l}$,
is the spin-orbit scattering strength for the corresponding $l$ channel, while the rest is described merely
in terms of the nonrelativistic phase shifts. Figure~\ref{fig.:SOC} shows the two discussed parts separately.
In contrast to the SOC part related to $l=1$, which has just a small enhancement going from Lu to Au,
the corresponding part for $l=2$ shows the resonance behavior~\cite{Friedel58,Daniel65} caused by
the $5d$ shell filling (see Table S2 in the Supplemental Material~\cite{Supplementary}).
This part is enhanced around W and reduces towards both sides. As a result, in the case of Au impurities
the spin-orbit scattering strength in the $p$ channel is four times larger than for $d$ states.
However, from Hf to Ir both contributions are of comparable magnitude and how they add up is determined by
the second part of the corresponding $C_{l l'}$ terms, which is given by
\begin{equation}\label{eq.:trigonometry}
\begin{array}{ll}
\sin{(2 \eta_l - \eta_{l'})} \sin{\eta_{l'}} \\ \qquad\quad
=  2\sin{\eta_l}\sin{\eta_{l'}}\cos{(\eta_l-\eta_{l'})} - \sin^2{\eta_{l'}}\ .
\end{array}
\end{equation}
Here, the used identity helps us to highlight a further important point, since the first term on the r.h.s of
Eq.~(\ref{eq.:trigonometry}) is the same for $l l'$ and $l' l$ contributions. The difference is caused solely
by the second term, which depends on $\eta_2$ and $\eta_1$ for the $p$$-$$d$ and $d$$-$$p$ contributons, respectively.
Thus, the nature of the scattering-in term, which is responsible for the structure of Eq.~(\ref{eq.:sigma_skew}),
leads to the coupling of the spin-orbit scattering in the $p$ channel with the potential scattering in the $d$ channel.
For the considered systems, the latter one is enhanced by the resonance properties of the $5d$ impurities in copper.
In contrast to the results of Ref.~\onlinecite{Fert11}, the subtle combination of the two scattering mechanisms causes
a positive SHC, and consequently a positive SHA, for all the considered systems.

Moreover, the main trend of the total SHC with a significant enhancement for Pt and Au, as seen in
Figs.~\ref{fig.:SigmaYX_total} and \ref{fig.:SigmaYX_ll}, is entirely provided by the behavior of $\tau_0$.
This is illustrated by Fig.~4 where we show the momentum relaxation time in addition to the two dominant
contributions in the braces of Eq.~(\ref{eq.:SHC_final}). With this expression, the strong correlation between
$\tau_0^2$ and the total SHC of Figs.~\ref{fig.:SigmaYX_total} and \ref{fig.:SigmaYX_ll} points to the crucial
role of the potential scattering strength by itself for the magnitude of the SHC.

In summary, we developed a phase shift model for the skew-scattering mechanism of the SHE as an extension
of the resonant scattering model of Ref.~\onlinecite{Fert11}. Exploiting this approach we show that the spin
Hall conductivity results from a subtle interplay between the spin-orbit and potential scattering in different
$l$ channels, caused by the structure of the scattering-in term (vertex corrections). For $5d$ impurities
in copper this leads to a crucial importance of the contribution related to the spin-orbit coupling of $p$ states,
which explains the unanticipated behavior of the spin Hall conductivity. The proposed phase shift model provides
good agreement with results of first-principles approaches and highlights the complexity of the specific problem.
In combination with parameters derived from \emph{ab initio} calculations this model can be used for a simplified
description of the SHE in dilute alloys based on noble metals. However, we emphasize the utmost importance to carefully
consider all leading terms.

\begin{figure}[h]
\includegraphics[width=0.85\LL]{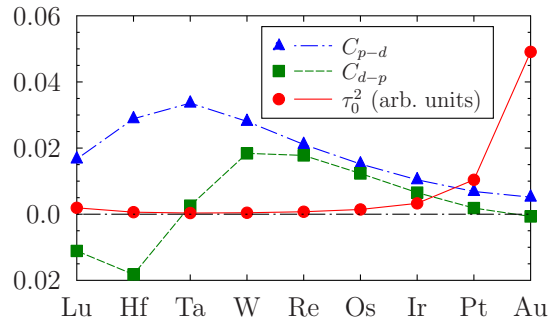}
\caption{(Color online) The contributions
$C_{p-d}$ and $C_{d-p}$ are shown versus $\tau_0^2$ (in arbitrary units).}
\label{fig.:Tau}
\end{figure}

The work was partially supported by the Deutsche Forschungsgemeinschaft (DFG) via SFB~762.
The authors K.C., H.E. and D.K. acknowledge support from the DFG via SFB~689 and SPP~1538.
In addition, M.G. acknowledges financial support from the DFG via a research fellowship
(GR3838/1-1).

\end{document}